\pdfoutput=1
\documentclass[%
prd,amsmath,aps,floats,amssymb, floatfix, superscriptaddress,nofootinbib,twocolumn,preprintnumbers
]{revtex4-2}

\usepackage{graphicx}
\usepackage{dcolumn}
\usepackage{bm}
\usepackage{cprotect}

\newcommand{\beq}{\begin{equation}}
\newcommand{\beqa}{\begin{eqnarray}}
\newcommand{\eeq}{\end{equation}}
\newcommand{\eeqa}{\end{eqnarray}}

\newcommand{\bfn}{\mathbf{n}}
\newcommand{\bfl}{\mathbf{l}}

\newcommand{\simgt}{\lower.5ex\hbox{$\; \buildrel > \over \sim \;$}}
\newcommand{\simlt}{\lower.5ex\hbox{$\; \buildrel < \over \sim \;$}}

\newcommand{\bd}[1]{\mbox{\boldmath $#1$}}

\usepackage{xcolor}

\usepackage{amsmath,amssymb,natbib,latexsym,times}

\begin{document}


\title{
Constraints on Annihilating Dark Matter from Gamma-Ray Background–Galaxy Shape Correlations: Model-independent Null Results and Moderate Template-based Signals
}

\author{Masato Shirasaki}
\email{masato.shirasaki@nao.ac.jp}
\affiliation{National Astronomical Observatory of Japan (NAOJ), National Institutes of Natural Sciences, Tokyo 181-8588, Japan}
\affiliation{The Institute of Statistical Mathematics, Tokyo 190-8562, Japan}
\affiliation{RIKEN Center for Advanced Intelligence Project, Tokyo 103-0027, Japan}

\author{Deheng Song}%
\affiliation{Universit{\'e} Libre de Bruxelles, Science Faculty CP230, B-1050 Brussels, Belgium}

\author{Oscar Macias}
\affiliation{Department of Physics and Astronomy, San Francisco State University, San Francisco, CA 94132, USA}
\affiliation{GRAPPA - Gravitational and Astroparticle Physics Amsterdam,
University of Amsterdam, Science Park 904, 1098 XH Amsterdam, The Netherlands}

\author{Shunsaku Horiuchi}
\affiliation{Department of Physics, Institute of Science Tokyo,
2-12-1 Ookayama, Meguro-ku, Tokyo 152-8551, Japan}
\affiliation{Kavli Institute for the Physics and Mathematics of the Universe, 5-1-5 Kashiwanoha Kashiwa, Chiba 277-8583, Japan}
\affiliation{Center for Neutrino Physics, Department of Physics, Virginia Tech, Blacksburg, VA 24061, USA}

\author{Naoki Yoshida}
\affiliation{RIKEN Center for Advanced Intelligence Project, Tokyo 103-0027, Japan}
\affiliation{Kavli Institute for the Physics and Mathematics of the Universe, 5-1-5 Kashiwanoha Kashiwa, Chiba 277-8583, Japan}
\affiliation{Department of Physics, School of Science, The University of Tokyo, 7-3-1 Hongo, Bunkyo, Tokyo 113-0033, Japan}

\date{\today}

\begin{abstract}
We revisit the cross-correlation between the unresolved $\gamma$-ray background and galaxy shapes to constrain the annihilation cross section of particle dark matter. 
Our analysis uses $\gamma$-ray photons from 14 years of observations with the Fermi Large Area Telescope (LAT), together with galaxy shape catalogs from the Dark Energy Survey Year 3 (DES Y3) and the Dark Energy Camera All Data Everywhere (DECADE) project, 
enabling us to probe cosmological large-scale signals over a common sky area of $\sim 12{,}000\,\mathrm{deg}^2$ shared by the $\gamma$-ray and galaxy data sets.
In order to better access signals from large-scale structure, we employ a Fourier-space estimator for the cross-correlation in contrast to the previous DES Y3 analysis. 
We find that our measurements are consistent with a null detection in a model-independent $\chi^2$ test, while template-based analyses yield signals at the $\sim 3\sigma$ level. 
Our null results exclude an enhanced annihilation cross section for wino-like dark matter with a mass of $2$--$3$ TeV under a modest substructure boost factor of $\sim 30$ in Milky Way-sized halos. For larger boost factors of $\sim 100$, the constraints become significantly stronger and exclude the canonical thermal annihilation cross section $\langle \sigma v \rangle = 3 \times 10^{-26}\,\mathrm{cm}^3/\mathrm{s}$ for a $7$--$40$ GeV dark matter particle annihilating into $b\bar{b}$ or $\tau^{+}\tau^{-}$.
The template-based analysis favors a power-law $\gamma$-ray energy dependence of 
the cross-correlation, but also indicates deviations from that expected based on the mean intensity of the unresolved $\gamma$-ray background around 100 GeV. 
We further consider decaying dark matter scenarios and derive $2\sigma$ lower limits on the particle lifetime of $\sim 10^{26}$--$10^{27}\,\mathrm{s}$, depending on the decay channel.
\end{abstract}
\maketitle


\section{Introduction}

The nature of cosmic dark matter (DM) remains one of the greatest mysteries in modern physics and astronomy. Weakly interacting massive particles (WIMPs) are well-motivated candidates that can naturally explain the observed DM abundance if their mass lies between 10 GeV and 10 TeV and their self-annihilation occurs at the weak scale \cite{Jungman:1995df}. A key prediction of the WIMP scenario is a spatial cross-correlation between the products of WIMP annihilations and large-scale structure. This is because WIMPs trace the matter distribution and their annihilation rate is enhanced in high-density halos corresponding to nodes of the cosmic web \cite{Fornengo:2013rga}.

The matter distribution can be probed directly by gravitational lensing. Foreground gravitational potentials distort the shapes of background galaxies, producing coherent alignments over large angular scales. This effect, known as cosmic shear, traces the matter distribution in an unbiased way. In this paper, we study cross-correlations between the unresolved extragalactic $\gamma$-ray background and cosmic shear, assuming a WIMP model that annihilates and produces high-energy photons.

Beyond its connection to DM, this cross-correlation also offers a unique probe of high-energy astrophysical processes. The unresolved $\gamma$-ray background is expected to arise from a mixture of sources, including active galactic nuclei, star-forming galaxies, and potentially DM annihilation or decay \cite{Fornasa:2015qua}. By correlating $\gamma$ rays with tracers of large-scale structure, one can statistically disentangle these contributions through their different redshift and environmental dependencies. In this context, weak-lensing measurements are particularly powerful, as they directly trace the matter distribution and enable a clean separation between astrophysical and dark matter components \cite{Camera:2014rja}. With the rapid growth of both $\gamma$-ray datasets and wide-field imaging surveys, such analyses have become an increasingly powerful tool for both cosmology and high-energy astrophysics.

Following the first theoretical prediction in Ref.~\cite{Camera:2012cj}, observational searches for this signal have progressed steadily. The first measurement using the Canada-France-Hawaii Lensing Survey reported no detection due to limited sky coverage and large statistical uncertainties \cite{Shirasaki:2014noa}. Subsequent studies improved sensitivity by expanding survey areas \cite{Shirasaki:2016kol, Troster:2016sgf, Shirasaki:2018dkz, DES:2019ucp, DES:2025ulp, Zhang:2026ysp}, but the detection status remains unsettled. Ref.~\cite{DES:2019ucp} reported the first detection using Dark Energy Survey (DES) Year 1 data and 9 years of Fermi Large Area Telescope (LAT) observations, based on real-space correlations over 
about $1500\,\mathrm{deg}^2$. A matched-filter analysis with halo-model-inspired templates yielded a signal-to-noise ratio of 5.3, later improved to 8.9 with updated datasets, though still relying on model assumptions \cite{DES:2025ulp}. In contrast, Ref.~\cite{Zhang:2026ysp} analyzed Fourier-space correlations using 15-year Fermi LAT data and Kilo-Degree Survey measurements over $\sim1000\,\mathrm{deg}^2$, finding results consistent with null using a model-independent test.

Motivated by these developments, we revisit the cross-correlation between the unresolved $\gamma$-ray background and galaxy shapes using DES Year-3 (Y3) data. We also include the extended galaxy-shape catalogs from the DECam All Data Everywhere (DECADE) project, enabling the largest sky coverage to date. Unlike Ref.~\cite{DES:2025ulp}, we perform the analysis in Fourier space, where statistical uncertainties are quantified in a controlled manner. We divide the data into 9 energy and 4 redshift bins to study the dependence of the signal on $\gamma$-ray energy and source redshift. We evaluate detection significance using a conservative, model-independent $\chi^2$ test and verify robustness against systematic effects in shape measurements. We also perform a template-based analysis similar to Ref.~\cite{DES:2025ulp} to assess consistency with previous claims.

The paper is organized as follows. In Section~\ref{sec:data}, we describe the $\gamma$-ray and galaxy data and outline the analysis method. Section~\ref{sec:model} introduces our theoretical models. In Section~\ref{sec:results}, we present the results, derive constraints on annihilating DM, and discuss detection significance using different templates. We conclude in Section~\ref{sec:conc}. Throughout, we adopt a flat $\Lambda$CDM cosmological model with the parameters $H_0 = 100h\, \mathrm{km}/\mathrm{s}/\mathrm{Mpc}$ with $h = 0.68$, $\Omega_\mathrm{m0} = 0.315$, $\Omega_\Lambda = 0.685$, and $\sigma_8 = 0.83$.

\section{Data}\label{sec:data}

\begin{table*}
\caption{\label{tab:basics_gray}
This table lists the $\gamma$-ray energy bins, the mean intensity $\bar{I}_\gamma$, and the sky fraction $f_\mathrm{sky}$ of the data region. The mean intensity is computed after subtracting Galactic $\gamma$-ray emission. The data region is defined by the combined masks of the $\gamma$-ray, DES Y3, and DECADE datasets.}
\begin{ruledtabular}
\begin{tabular}{lccccccccc}
$E_\gamma$ (GeV)
& $0.63$--$1.25$ & $1.25$--$2.51$ & $2.51$--$5.01$ & $5.01$--$10.0$
& $10.0$--$19.9$ & $19.9$--$39.8$ & $39.8$--$79.4$ & $79.4$--$158$ & $158$--$1000$ \\
 $\bar{I}_\gamma$\footnote{In units of $\mathrm{GeV}^{-1}\mathrm{cm}^{-2}\mathrm{s}^{-1}\mathrm{sr}^{-1}$.} 
 & $1.10\times10^{-6}$ & $2.57\times10^{-7}$ & $6.93\times10^{-8}$ & $1.88\times10^{-8}$ &$4.74\times10^{-9}$ & $1.13\times10^{-9}$ & $2.56\times10^{-10}$ & $5.49\times10^{-11}$ & $2.60\times10^{-12}$ \\
 $f_\mathrm{sky}$ 
 & 0.019 & 0.12 & 0.21 & 0.27 & 0.29 & 0.29 & 0.29 & 0.30 & 0.30 \\
\end{tabular}
\end{ruledtabular}
\end{table*}

\subsection{Fermi LAT}

The $\gamma$-ray data used in this analysis are based on 14 years of Fermi-LAT Pass8 observations, spanning mission elapsed times 239557417–681264888, corresponding to the period from 04 August 2008 to 04 August 2022. We select events in the energy range $100\mathrm{MeV}$–$1~\mathrm{TeV}$ using the \texttt{P8R3\_SOURCEVETO\_V3} instrument response functions and event type 56, which corresponds to the combined PSF1, PSF2, and PSF3 event classes. This selection excludes PSF0, the quartile with the poorest directional reconstruction. An all-sky event selection is performed with a zenith-angle cut of $z<90^\circ$ to suppress contamination from the Earth limb. Standard good-time intervals are imposed by requiring \texttt{DATA\_QUAL>0} and \texttt{LAT\_CONFIG==1}. The selected events are binned into Galactic-coordinate \texttt{HEALPix} maps with $\mathrm{NSIDE}=512$ and 40 logarithmically spaced energy bins spanning 100 MeV to 1 TeV.

We extract the unresolved $\gamma$-ray background component from the selected LAT maps by subtracting the dominant Galactic foreground and resolved-source contributions. The analysis is restricted to the high-latitude region $|b|\geq30^\circ$ in order to exclude the Galactic plane. In addition, we apply an energy-dependent mask to remove emission from resolved sources in the 14-year 4FGL-DR4 catalog. For each source, the LAT point-spread function is computed at the source position using \texttt{gtpsf}, with the same livetime cube, instrument response functions, and event type as used in the data selection. In each energy bin, the corresponding PSF containment angle is evaluated for a containment fraction of 95\%. A \texttt{HEALPix} pixel is masked if its angular separation from any cataloged source is smaller than the corresponding containment angle. Applying this procedure to all cataloged sources yields the energy-dependent resolved-source mask.

The unresolved $\gamma$-ray background is then obtained through a template-based decomposition of the selected LAT counts maps. For each model component, we construct response-convolved templates using the same data selection criteria. The Galactic foreground is modeled using the \texttt{gll\_iem\_v07} interstellar emission model, while the resolved-source contribution is described by the 4FGL-DR4 source template \texttt{gll\_psc\_v35.fit}. The unresolved $\gamma$-ray background is approximated by the residual emission after subtracting the Galactic diffuse and 4FGL-DR4 source models from the masked data. The residual counts are converted to intensity by dividing by the exposure and pixel solid angle.

\begin{figure}[!t]
\includegraphics[clip, width=1.\columnwidth]{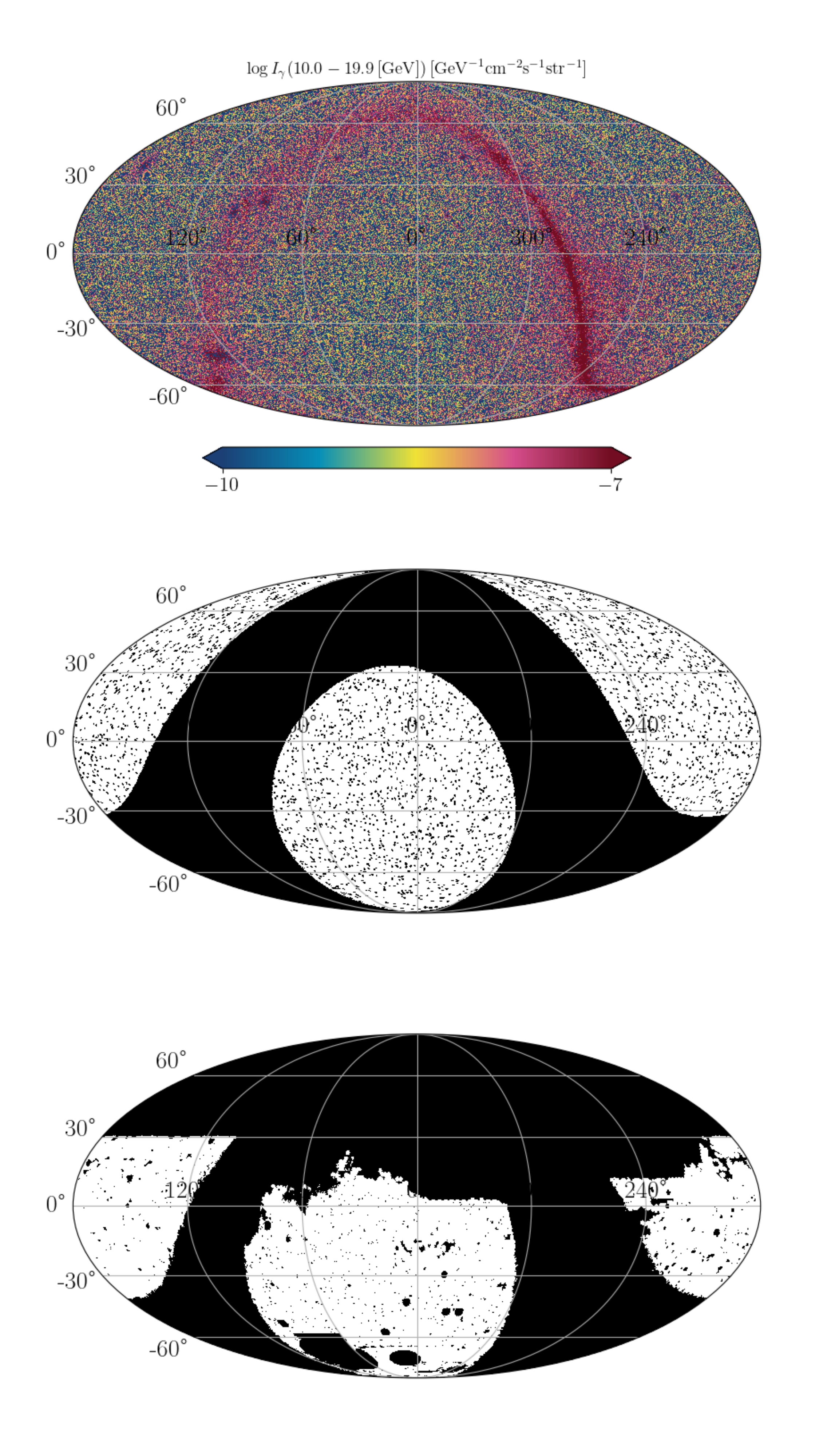}
\cprotect\caption{\label{fig:FOV} 
The field of view of our datasets. The top panel shows the observed $\gamma$-ray intensity in units of $\log [\mathrm{GeV}^{-1}\mathrm{cm}^{-2}\mathrm{s}^{-1}\mathrm{str}^{-1}]$ in the energy range of $10.0$–$19.9\, \mathrm{GeV}$. The middle panel shows the same energy range, with the black region indicating the mask applied to the $\gamma$-ray map, including a conservative cut around the Galactic plane and masked resolved sources. The bottom panel shows the field of view of the shear maps (white region) after combining the DES Y3 and DECADE data. We use the equatorial coordinate system throughout this figure.
}
\end{figure}

To construct the final flux maps, we follow a procedure similar to that of Ref.~\cite{DES:2025ulp}. 
Starting from residual count maps in 40 logarithmically spaced micro-energy bins spanning 100 MeV to 1 TeV, 
we divide each map by the corresponding exposure map
computed in the same micro bin and further correct for the pixel area, yielding flux maps in units of $\mathrm{photons}/\mathrm{cm}^{2}/\mathrm{s}/\mathrm{sr}$. 
These micro-bin flux maps are then summed into nine broader (macro) energy bins covering 630 MeV to 1 TeV; the resulting maps are subsequently normalized by the corresponding macro energy bin widths, as listed in Table~\ref{tab:basics_gray}. 
We exclude energies below $0.63\,\mathrm{GeV}$ due to the insufficient angular resolution at low energies for our analysis. 
A summary of the resulting $\gamma$-ray data sets is provided in Table~\ref{tab:basics_gray}. 
The top and middle panels of Fig.~\ref{fig:FOV} show the observed $\gamma$-ray flux in the energy range $10.0$–$19.9\,\mathrm{GeV}$ and the corresponding masked regions, respectively.

\subsection{DES Y3 and DECADE}

We use weak lensing shear catalogs from the DES Y3 \cite{DES:2020ekd} and the DECADE project \cite{Anbajagane:2025hlf, Anbajagane:2025goh}. These contain approximately 100 million and 170 million galaxies, respectively, and together cover 13,000 square degrees. Both rely on DECam imaging. DES Y3 was designed as a dedicated survey and provides a more homogeneous dataset. In contrast, DECADE combines public DECam imaging, mainly from structured wide-area programs (DELVE, DECaLS, DeROSITAS) optimized for near-uniform coverage, along with smaller programs. This leads to greater variation in exposure time and image quality \cite{Anbajagane:2025oww}.
In both datasets, the {\tt Metacalibration} method has been applied to estimate shear components \cite{Huff:2017qxu, Sheldon:2017szh}. The catalogs are split into four tomographic bins based on source photometric redshifts to probe the evolution of large-scale structure \cite{DES:2020ebm, Anbajagane:2025pvn}.
The average source redshifts at individual bins are given by 0.339, 0.528, 0.752, and 0.952, respectively. 

We construct shear maps by pixelizing the catalogs into the two shear components using {\tt HEALPix} \cite{Gorski:2004by} with $\mathrm{NSIDE}=512$, corresponding to 6.87 arcmin pixels. The shear in each pixel is given by
\beqa
\gamma^{\nu}_\mathrm{obs} = \frac{\sum_{j=1}^{n} \epsilon^{\nu} w_j}{\bar{R}\sum_{j=1}^{n} w_j}, \label{eq:shear_obs_est}
\eeqa
where $\nu$ indexes the shear components, $n$ is the number of galaxies, $w_j$ is the inverse-variance weight per galaxy \cite{DES:2020ekd}, and $\bar{R}$ is the mean {\tt Metacalibration} response. We apply Eq.~(\ref{eq:shear_obs_est}) separately to DES Y3 and DECADE. We subtract any non-zero mean shear before map construction.
The white region of the bottom panel in Figure~\ref{fig:FOV} shows the field of view in our shear maps.
    
\subsection{Power spectra and covariance}

We estimate all power spectra using the pseudo-$C_\ell$ algorithm \cite{Hivon:2001jp} implemented in {\tt NaMaster} \cite{Alonso:2018jzx}. This method analytically accounts for the coupling between multipoles induced by the sky mask, encoded in the mode-coupling matrix. Details of the method, including its extension to spin-2 fields such as cosmic shear, are given in Ref.~\cite{Alonso:2018jzx}.

We compute the covariance matrix analytically. 
We evaluate the disconnected Gaussian term using the improved Narrow Kernel Approximation \cite{Nicola:2020lhi, Garcia-Garcia:2019bku}, 
including mode-coupling effects through approximate pseudo-$C_\ell$ methods.
We do not consider potential non-Gaussian covariances as the previous DES Y3 analysis validates the Gaussian-covariance approximation \cite{DES:2025ulp}.

For the power spectrum measurement, we adopt a hybrid binning scheme following Ref.~\cite{LaPosta:2024kls}. We use linear bins with width $\Delta \ell = 30$ for $\ell < 240$, and logarithmic bins with $\Delta \log_{10} \ell = 0.055$ for $30 < \ell < 3\mathrm{NSIDE}$. We restrict the analysis to $30 < \ell < 512$ to avoid pixelization artifacts and potential B-mode contaminations\footnote{Since B-mode signals in our analysis arise from systematic effects such as calibration errors in the point spread function in galaxy image measurements, as well as inaccuracies in mask deconvolution in Fourier-space analyses, it is crucial to suppress and control them to ensure a robust and unbiased measurement.}.
This results in 13 bandpowers per $C_\ell$.
Note that we measure the power spectra in units of $\mathrm{GeV}^{-1}\mathrm{cm}^{-2}\mathrm{s}^{-1}$.
Because we have nine $\gamma$-ray maps (i.e., nine macro energy bins) and twelve galaxy shape catalogs 
(i.e., four different source redshift bins and three separate survey windows),
$9\times 12 = 108$ power spectra are available in total.
In practical anlyses, we do not consider cross-survey covariances between different shape catalogs.

To correct for the point-spread function (PSF) of the $\gamma$-ray observations, we derive the PSF in Fourier space at each energy bin following Ref.~\cite{Shirasaki:2016kol}, and divide the measured power spectrum by the corresponding PSF.

\section{Model}\label{sec:model}

In this section, we summarize two classes of cross power spectra 
between the unresolved $\gamma$-ray background and cosmic shear. 
The first assumes that DM particles produce $\gamma$-ray photons through annihilation and simultaneously trace the large-scale structures probed by weak-lensing surveys. The second are two phenomenological models developed in Ref.~\cite{DES:2025ulp} to reproduce their cross-correlation signals.

\subsection{Observables}

\subsubsection*{Gamma-ray intensity}
The cumulative $\gamma$-ray intensity $I_\gamma$ along the direction $\hat{\bfn}$ is given by
\beqa
E_\gamma I_\gamma(\hat{\bfn}) =
\frac{c}{4\pi} \int {\rm d}z \frac{P_\gamma (E'_\gamma,\chi\hat{\bfn},z)}{H(z)(1+z)^4} e^{-\tau(E'_\gamma,z)}, \label{eq:Intensity}
\eeqa
where $E_\gamma$ is the observed energy and $E'_\gamma = (1+z) E_\gamma$ is the emitted energy at redshift $z$. The Hubble parameter is $H(z) = H_0 [\Omega_{\rm m0}(1+z)^3+\Omega_\Lambda]^{1/2}$ in a flat Universe, and $\chi$ is the comoving distance to $z$. 
The attenuation factor $e^{-\tau}$ accounts for $\gamma$-ray absorption via pair production on background photons. We adopt $\tau(E'_\gamma, z)$ from Ref.~\citep{Gilmore:2011ks}.
The volume emissivity $P_\gamma$, defined as the emitted photon energy per unit volume, time, and energy, is
\beqa
P_\gamma(E_\gamma, \chi\hat{\bfn}, z)=
E_\gamma {\cal S}(E_\gamma, z)
{\cal F}(\chi\hat{\bfn}, z), \label{eq:Emissivity}
\eeqa
where ${\cal S}$ is the source function and ${\cal F}$ is the source density field. Substituting Eq.~(\ref{eq:Emissivity}) into Eq.~(\ref{eq:Intensity}) gives
\beqa
I_\gamma(\bfn) &=& \int \mathrm{d}\chi \, W_{\gamma}(z)\, {\cal F}(\chi\bfn, z), \\
W_{\gamma}(z) &=& \frac{{\cal S}(E'_\gamma, z)e^{-\tau(E'\gamma,z)}}{4\pi (1+z)^3}.
\eeqa

\subsubsection*{Weak lensing}
Weak lensing is described by the distortion matrix
\beqa
A_{ij} = \frac{\partial \beta^{i}}{\partial \theta^{j}}
\equiv \left(
\begin{array}{cc}
1-\kappa -\gamma_{+} & -\gamma_{\times}  \\
-\gamma_{\times} & 1-\kappa+\gamma_{+} \\
\end{array}
\right), \label{distortion_tensor}
\eeqa
where $\bd{\theta}$ and $\bd{\beta}$ denote observed and true positions, $\kappa$ is the convergence, and $\gamma_{(+,\times)}$ are the shear components. In the weak regime ($\kappa, \gamma \ll 1$), these quantities relate to second derivatives of the gravitational potential $\Phi$ \cite{Bartelmann:1999yn}.
Using the Poisson equation and the Born approximation, the shear field is
\beqa
\gamma_{(+,\times)}(\hat{\bfn}) &=&
\partial^{-2} \int_{0}^{\chi_{H}} {\rm d}\chi \, W_{\mathrm{WL}}(\chi) \nonumber \\
&&
\quad (\partial^2_x - \partial^2_y, 2 \partial_x \partial_y) \delta_{\rm m}(\chi\hat{\bfn},\chi), \label{eq:shear_delta}
\eeqa
where derivatives act on angular coordinates, $\partial^{-2}$ is the inverse Laplacian, and $\delta_{\rm m}$ is the matter overdensity. The lensing kernel is
\beq
W_{\mathrm{WL}}(\chi) = \frac{3}{2} \,\Omega_{\rm m0}H^2_{0}\, (1+z(\chi)) \int_{\chi}^{\chi_{H}} {\rm d}\chi^{\prime} p(\chi^{\prime})\frac{\chi(\chi^{\prime}-\chi)}{\chi^{\prime}}, \label{eq:lens_kernel}
\eeq
where $p(\chi)$ represents the normalized source redshift distribution, i.e.~$\int_{0}^{\chi_H} {\rm d}\chi \, p(\chi) =1$.
In practice, we use the publicly available redshift distributions of DES Y3 and DECADE catalogs to set the lensing kernel $W_{\mathrm{WL}}$.

\subsubsection*{Cross-correlations}
We characterize the cross-correlation between the $\gamma$-ray background and galaxy shapes in Fourier space. The galaxy shape is a spin-2 field, so the E/B-mode decomposition provides a clear interpretation. The E- and B-modes are defined as
\beqa
\gamma^\mathrm{obs}_{E}(\bfl) &=&
\gamma^\mathrm{obs}_{+}(\bfl) \cos 2\phi_{\bfl} + \gamma^\mathrm{obs}_{\times}(\bfl) \sin 2\phi_{\bfl}, \label{eq:Emode} \\
\gamma^\mathrm{obs}_{B}(\bfl) &=&
-\gamma^\mathrm{obs}_{+}(\bfl) \sin 2\phi_{\bfl} + \gamma^\mathrm{obs}_{\times}(\bfl) \cos 2\phi_{\bfl}, \label{eq:Bmode}
\eeqa
where $A(\bfl)$ is the Fourier transform of $A(\hat{\bfn})$ and $\phi_{\bfl} \equiv \tan^{-1}(l_2/l_1)$.
The observed shear field is
\beqa
\gamma^\mathrm{obs}_{(+,\times)}(\hat{\bfn}) = \gamma_{(+,\times)}(\hat{\bfn}) 
+ \gamma^{I}_{(+,\times)}(\hat{\bfn}) + \gamma^{N}_{(+,\times)}(\hat{\bfn}), \label{eq:obs_shear}
\eeqa
where the terms represent lensing shear, intrinsic alignment, and shape noise. Gravitational lensing produces only E-modes, so the B-mode vanishes in Eq.~(\ref{eq:shear_delta}).
Similarly, the observed $\gamma$-ray intensity is
\beqa
I_{\gamma, \mathrm{obs}}(\hat{\bfn}) =
I_{\gamma, \mathrm{LSS}}(\hat{\bfn}) + I_{\gamma, N}(\hat{\bfn}), \label{eq:obs_gammaray}
\eeqa
where $I_{\gamma, \mathrm{LSS}}$ is the extragalactic signal and $I_{\gamma, N}$ includes photon noise and Galactic emission.
The cross-correlation in Fourier space is defined as
\beqa
\langle A(\bfl) B(\bfl')\rangle \equiv (2\pi)^2 C^{AB}(\ell), \delta_\mathrm{D}(\bfl+\bfl').
\eeqa
Assuming no correlation between noise terms, the cross power spectrum becomes
\beqa
C^{\gamma E}(\ell) = \int \frac{\mathrm{d}\chi}{\chi^2} W_{\gamma}(z) W_{\mathrm{WL}}(z) P_{m{\cal F}}\left(k=\frac{\ell}{\chi}, z\right), \label{eq:cross_powerspec}
\eeqa
where $P_{m{\cal F}}$ is the three-dimensional power spectrum of ${\cal F}$ and $\delta_\mathrm{m}$. We adopt the Limber approximation \cite{Limber:1954zz} and neglect correlations between $I_{\gamma,\mathrm{LSS}}$ and intrinsic alignments $\gamma^{I}$. The latter is expected to be subdominant in current weak-lensing data \cite{Anbajagane:2025goh}.

\subsection{Annihilating Dark Matter}\label{subsec:DMann}

We follow Refs.~\cite{Shirasaki:2014noa, Shirasaki:2016kol} to model the $\gamma$-ray signal from DM annihilation. In Eq.~(\ref{eq:Emissivity}), the source function ${\cal S}$ and density field ${\cal F}$ are
\beqa
{\cal S} (E_\gamma, z) &=&
\frac{\langle \sigma v \rangle}{2 m^{2}_{\rm dm}}
\left[\frac{{\rm d}N_{\gamma, a}}{{\rm d}E_\gamma}+Q_{{\rm IC},a}(E_\gamma, z)\right], \\
{\cal F}(\chi \bfn, z) &=&
\left[\rho_{\rm dm}(\chi\bfn, z)\right]^{2},
\eeqa
where ${\rm d}N_{\gamma,a}/{\rm d}E_\gamma$ is the primary $\gamma$-ray spectrum per annihilation, $Q_{{\rm IC},a}$ is the inverse-Compton (IC) contribution, $\langle \sigma v \rangle$ is the velocity-averaged cross section, and $m_{\rm dm}$ is the DM particle mass.

We assume identical spatial distributions for primary and secondary photons. This is justified because $\sim$100 GeV electrons lose energy via IC scattering faster than they diffuse in halos \cite{Ishiwata:2009dk,Brunetti:2014gsa}, so propagation effects are negligible.

We consider benchmark annihilation channels with $100\%$
branching ratios into $b\bar{b}$, $\tau^+\tau^-$, $\mu^{+}\mu^{-}$, and $W^{+}W^{-}$. 
The primary spectra are computed using {\tt PPPC4DMID} \cite{Cirelli:2010xx}. Secondary IC emission from up-scattered background photons is included following Refs.~\cite{Ando:2015qda, Ando:2016ang}. We assume the cosmic microwave background as the background photon field and neglect the extragalactic background light, which contributes at the $\sim5\%$ level at $z=0$ and less at higher redshift \cite{Ando:2016ang}.

The cross power spectrum in Eq.~(\ref{eq:cross_powerspec}) involves $P_{m{\cal F}} = \bar{\rho}_{\rm dm}^2 P_{\delta,\delta^2}$, where 
$P_{\delta,\delta^2}$ represents the cross power spectrum between matter overdensity field and DM overdensity squared, and $\bar{\rho}_{\rm dm}$ is the mean DM density. We compute $P_{\delta,\delta^2}$ using the halo model as in Refs.~\cite{Shirasaki:2014noa, Shirasaki:2016kol}. It consists of one-halo and two-halo terms, describing intra-halo correlations and halo clustering, respectively.
We adopt the halo mass function and bias from Refs.~\cite{Tinker:2008ff, Tinker:2010my}, assume NFW profiles \cite{Navarro:1996gj}, and use the concentration model of Ref.~\cite{Prada:2011jf}.
To account for DM substructures, we include a boost factor $b_\mathrm{sh}$ with three models: HIGH \cite{Gao:2011rf}, MID \cite{Moline:2016pbm}, and LOW \cite{Ando:2019xlm}. These span the uncertainty on small scales and correspond to typical values of $100$, $30$, and $3$ for $10^{12}\,M_\odot$ halos at $z=0$.

\begin{figure*}[!t]
\includegraphics[clip, width=2.\columnwidth]{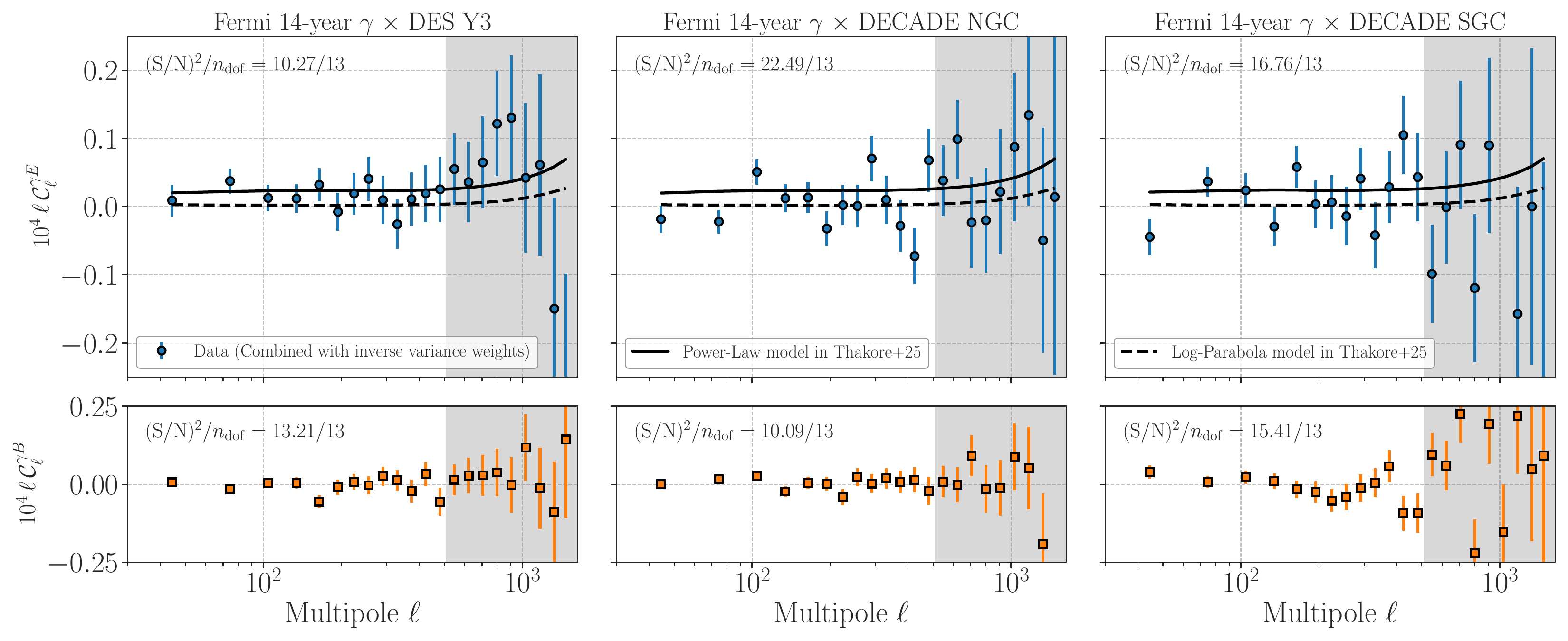}
\cprotect\caption{\label{fig:cl_stacked} 
A summary of our power-spectrum measurements.
We construct a composite power spectrum by combining measurements across energy and source redshift bins with inverse-variance weights.
The upper panels show E-mode spectra for different survey regions, and the lower panels show the corresponding B-mode spectra.
Each panel lists the signal-to-noise ratio and degrees of freedom, indicating no significant cross-correlation between the unresolved $\gamma$-ray background and galaxy shapes.
For reference, we include two phenomenological models from Ref.~\cite{DES:2025ulp}: the Power-law model (solid) and the Log-parabola model (dashed).
Data points in the gray regions are excluded to avoid potential artifacts in the power-spectrum measurements.
}
\end{figure*}

\subsection{Phenomenological models}\label{subsec:DESY3_model}

For a consistency test, we also consider two phenomenological models for the cross-correlation between the unresolved $\gamma$-ray background and cosmic shear, following Ref.~\cite{DES:2025ulp}. These models include two components motivated by the halo model:
\beqa
\xi (\theta, E_\gamma, \bar{z}_\mathrm{source}) &=& f_\mathrm{1h}(E_\gamma, \bar{z}_\mathrm{source}) \xi_\mathrm{1h}(\theta, E_\gamma) \nonumber \\
&&
+ f_\mathrm{2h}(E_\gamma, \bar{z}_\mathrm{source}) \xi_\mathrm{2h}(\theta, E_\gamma),
\label{eq:DESY3_model}
\eeqa
where $\xi$ is the real-space cross-correlation at angular separation $\theta$, $\gamma$-ray energy $E_\gamma$, and mean source redshift $\bar{z}_\mathrm{source}$.

Under the flat-sky approximation, Eq.~(\ref{eq:DESY3_model}) becomes
\beqa
C^{\gamma E}(\ell, E_\gamma, \bar{z}_\mathrm{source}) = \int \mathrm{d}^2\theta \, \xi (\theta, E_\gamma, \bar{z}_\mathrm{source}) J_2(\ell \theta),
\label{eq:Cell_DESY3model}
\eeqa
where $J_2(x)$ is the second-order Bessel function of the first kind.

Ref.~\cite{DES:2025ulp} considered two parameterizations of the prefactors $f_\mathrm{1h}$ and $f_\mathrm{2h}$. The first adopts power-law forms:
\beqa
f_\mathrm{1h} &=& A_1 \left(E_\gamma / E_0 \right)^{-\alpha_1}\left(\frac{1+\bar{z}_\mathrm{source}}{1+z_0}\right)^{\beta_1}, \label{eq:f1h_DESY3model} \\
f_\mathrm{2h} &=& A_2 \left(E_\gamma / E_0 \right)^{-\alpha_2}\left(\frac{1+\bar{z}_\mathrm{source}}{1+z_0}\right)^{\beta_2}, \label{eq:f2h_DESY3model}
\eeqa
with $E_0 = 13.7, \mathrm{GeV}$ and $z_0 = 0.64$. We adopt the best-fit parameters from Ref.~\cite{DES:2025ulp}: $A_1=17.3 \times 10^{-12}, \alpha_1=2.13, \beta_1=4.63, A_2 = 0.077, \alpha_2=2.01,$ and $\beta_2=4.83$.

The second log-parabola model introduces additional energy dependence by replacing $\alpha_i$ $(i=1,2)$ as
\beqa
\alpha_i \rightarrow \alpha_i - \gamma_i \log_{10}(E_\gamma/E_0),
\eeqa
with best-fit values $A_1=1.48 \times 10^{-12}, \alpha_1=0.871, \beta_1=5.17, \gamma_1 =0.073$, $A_2 =0.20, \alpha_2=1.94, \beta_2=4.83$, and $\gamma_2=1.61$ \cite{DES:2025ulp}.

For the angular templates $\xi_\mathrm{1h}(\theta, E_\gamma)$ and $\xi_\mathrm{2h}(\theta, E_\gamma)$, we neglect the $E_\gamma$ dependence and digitize their forms from the top-left panel of Figure~3 in Ref.~\cite{DES:2025ulp}\footnote{We use the web service of \url{https://plotdigitizer.com/}. Ref.~\cite{DES:2025ulp} does not provide details on the calculation of $\xi_\mathrm{2h}$.}.
We use these phenomenological models to test consistency with Ref.~\cite{DES:2025ulp}, which justifies this simplified treatment.

\section{Results}\label{sec:results}

\subsection{Measurements of power spectra}\label{subsec:measurements}

We present measurements of the power spectra of the unresolved $\gamma$-ray background and galaxy shapes, which we use to constrain annihilating DM.

Figure~\ref{fig:cl_stacked} summarizes the results for three survey regions: DES Y3, DECADE Northern Galactic Cap (NGC), and DECADE Southern Galactic Cap (SGC). The top panels 
show E-mode power spectra, while the bottom panels show B-modes, which indicate potential lensing systematics. 
For the figure, we define the weighted, dimensionless power spectrum as
\beqa\label{eq:combined_cl}
{\cal C}^{\gamma E}(\ell)
= \sum_{\alpha, \beta} {\cal W}_{\alpha\beta}(\ell) \,
C^{\gamma E}_{\alpha\beta}(\ell) / \bar{I}_{\gamma,\alpha},
\eeqa
where $C^{\gamma E}_{\alpha\beta}$ is the measurement in the $\alpha$-th energy and $\beta$-th redshift bin, and $\bar{I}_{\gamma, \alpha}$ is the mean intensity. The sum runs over all bins for a given catalog. We choose the weights ${\cal W}_{\alpha\beta}$ to minimize the variance:
\beqa
{\cal W}_{\alpha\beta} (\ell) =
\frac{{\rm Var}^{-1}[C^{\gamma E}_{\alpha\beta}](\ell)}
{\sum_{\alpha, \beta}{\rm Var}^{-1}[C^{\gamma E}_{\alpha\beta}](\ell)},
\eeqa
where ${\rm Var}[C^{\gamma E}_{\alpha\beta}]$ is the variance. 
We stack the B-mode spectra in the same way.

We quantify the signal using the signal-to-noise ratio of Eq.~(\ref{eq:combined_cl}):
\beqa
(\mathrm{S/N})^2 &=& \sum_{i,j}
{\cal C}^{\gamma E}(\ell_{i})
\Gamma^{-1}_{{\cal C}, ij}
{\cal C}^{\gamma E}(\ell_{j}), \\ \label{eq:chi2_combined}
\Gamma_{{\cal C}, ij} &=& \sum_{\alpha, \beta, \alpha', \beta'}
{\cal W}_{\alpha \beta}(\ell_{i}) / \bar{I}_{\gamma,\alpha}
{\cal W}_{\alpha' \beta'}(\ell_{j}) / \bar{I}_{\gamma,\alpha'} \nonumber \\
&&
\quad \quad \quad \quad\times\Gamma_{ij}(\alpha'\beta'\alpha'\beta'),
\eeqa
where $\Gamma_{ij}(\alpha'\beta'\alpha'\beta')$ 
is the covariance of $C^{\gamma E}_{\alpha\beta}$.

We find no significant detection in DES Y3 or DECADE. The $(\mathrm{S/N})^2$ per degree of freedom is $10.27/13$ for DES Y3, $22.49/13$ for DECADE NGC, and $16.76/13$ for DECADE SGC. The stacked B-modes are also consistent with null, indicating controlled systematics. We further examine individual energy and redshift bins in Appendix~\ref{apdx:individual_cl} and find no significant excess.

The black solid and dashed lines in the top panel of Fig.~\ref{fig:cl_stacked} show predictions from the DES-Y3 phenomenological models (Section~\ref{subsec:DESY3_model}). The simple power-law model provides a better fit. The detection significance of this model is discussed in Section~\ref{subsec:model_depend_s2n}.

\begin{figure*}[!t]
\includegraphics[clip, width=2.\columnwidth]{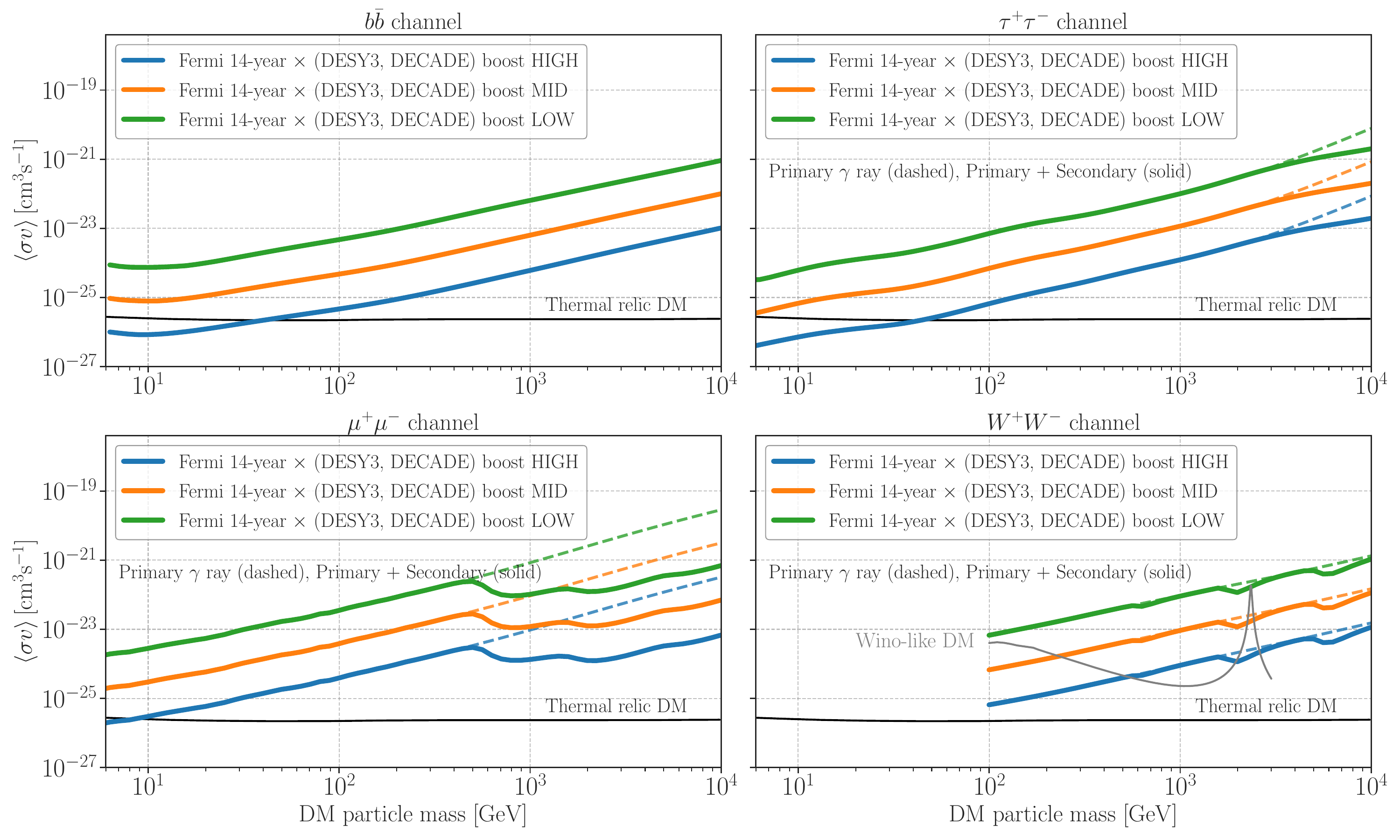}
\cprotect\caption{\label{fig:limit_DMann} 
95\% confidence upper limits on the DM annihilation cross section as a function of DM mass. The four panels show the $b\bar{b}$, $\tau^{+}\tau^{-}$, $\mu^{+}\mu^{-}$, and $W^{+}W^{-}$ channels. Solid (dashed) lines include (exclude) secondary $\gamma$ rays from inverse-Compton scattering. Colors indicate uncertainty from substructure modeling. The black horizontal line shows the canonical thermal relic cross section \cite{Steigman:2012nb}, and the gray line (in the bottom right panel) shows the prediction for wino DM \cite{Hryczuk:2011vi}.
}
\end{figure*}

\subsection{Likelihood analysis of DM annihilations}

Using the null detection of the cross-correlation, particle DM can be constrained via a maximum-likelihood analysis. We assume the data vector ${\bd D}$ follows a multivariate Gaussian distribution with covariance $\Gamma$. 
The $\chi^2$ statistic is
\beqa
\chi^2({\bd p}) = \sum_{i,j}\left[D_{i}-\mu_{i}({\bd p})\right]\Gamma^{-1}_{ij}\left[D_{j}-\mu_{j}({\bd p})\right], \label{eq:logL}
\eeqa
where ${\bd \mu}({\bd p})$ is the model prediction for parameters ${\bd p}$, computed as in Section~\ref{subsec:DMann}. The parameters are the DM mass $m_{\rm dm}$ and annihilation cross section $\langle \sigma v \rangle$.
The data vector ${\bd D}$ contains 108 measured power spectra. 
We derive 95\% confidence regions in the $(m_{\rm dm}, \langle \sigma v \rangle)$ plane using
\beqa
\Delta \chi^2({\bd p}) = \chi^2({\bd p})-\chi^2({\bd \mu}=0)=6.17.
\eeqa
As discussed in Section~\ref{subsec:DMann}, the boost factor $b_\mathrm{sh}$ strongly affects the model predictions. We therefore present results for three models of $b_\mathrm{sh}$ to bracket the modeling uncertainty.

Figure~\ref{fig:limit_DMann} shows the constraints for four annihilation channels: $b\bar{b}$, $\tau^{+}\tau^{-}$, $\mu^{+}\mu^{-}$, and $W^{+}W^{-}$. Dashed lines include only primary $\gamma$ rays, while solid lines include secondary emission from the IC scattering. The IC contribution manifests from relativistic $e^{\pm}$ upscattering ambient background photons and enhances the signal particularly for heavier DM, leading to significantly stronger constraints, typically by a factor of $\sim 10$. This effect is more pronounced for leptonic channels, which produce more energetic $e^{\pm}$.

As a particle-physics example, we consider wino-like DM annihilating into $W^{+}W^{-}$, 
which exhibits an enhanced cross section \cite{Hryczuk:2011vi}. The wino is a well-motivated candidate in anomaly-mediated supersymmetry breaking models \cite{Randall:1998uk, Giudice:1998xp}, consistent with the observed 125 GeV Higgs boson \cite{ATLAS:2012yve, CMS:2012qbp}, and naturally realized as the lightest supersymmetric particle \cite{Okada:1990vk,Ellis:1990nz,Haber:1990aw,Ellis:1991zd}. 
A wino with mass below $\sim 3$ TeV can reproduce the observed DM abundance \cite{Hisano:2006nn}. Figure~\ref{fig:limit_DMann} shows that our measurements exclude the enhanced annihilation cross section of wino-like DM with masses of 2–3 TeV, even for a moderate boost factor.

\begin{figure*}[!t]
\includegraphics[clip, width=2.\columnwidth]{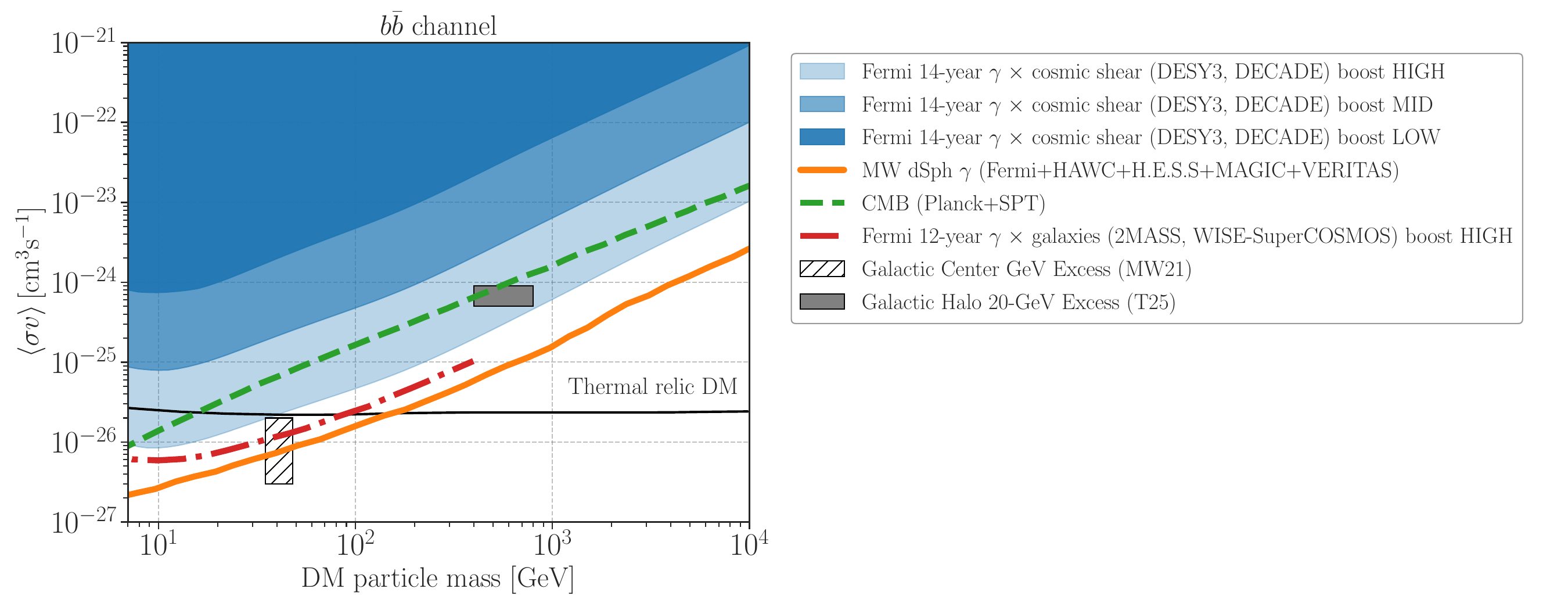}
\cprotect\caption{\label{fig:limit_DMann_comparisons} 
Upper limits on the DM annihilation cross section for the $b\bar{b}$ channel, compared with constraints from other probes.
The blue filled regions show our constraints, including uncertainties from DM substructure.
The orange line denotes limits from $\gamma$-ray observations of Milky Way dwarf spheroidal galaxies (dSph) \cite{Fermi-LAT:2025gei}.
The green dashed line shows constraints from the cosmic microwave background \cite{Myers:2025pfx}.
The red dash-dotted line indicates cosmological constraints from cross-correlations between galaxy number density and the unresolved $\gamma$-ray background under the HIGH boost-factor scenario \cite{Paopiamsap:2023uuo}.
The hatched and gray filled squares mark reported $\gamma$-ray excesses in the Galactic center \cite{DiMauro:2021qcf} and Galactic halo \cite{Totani:2025fxx}, respectively.
}
\end{figure*}

\begin{figure*}[!t]
\includegraphics[clip, width=1.8\columnwidth]{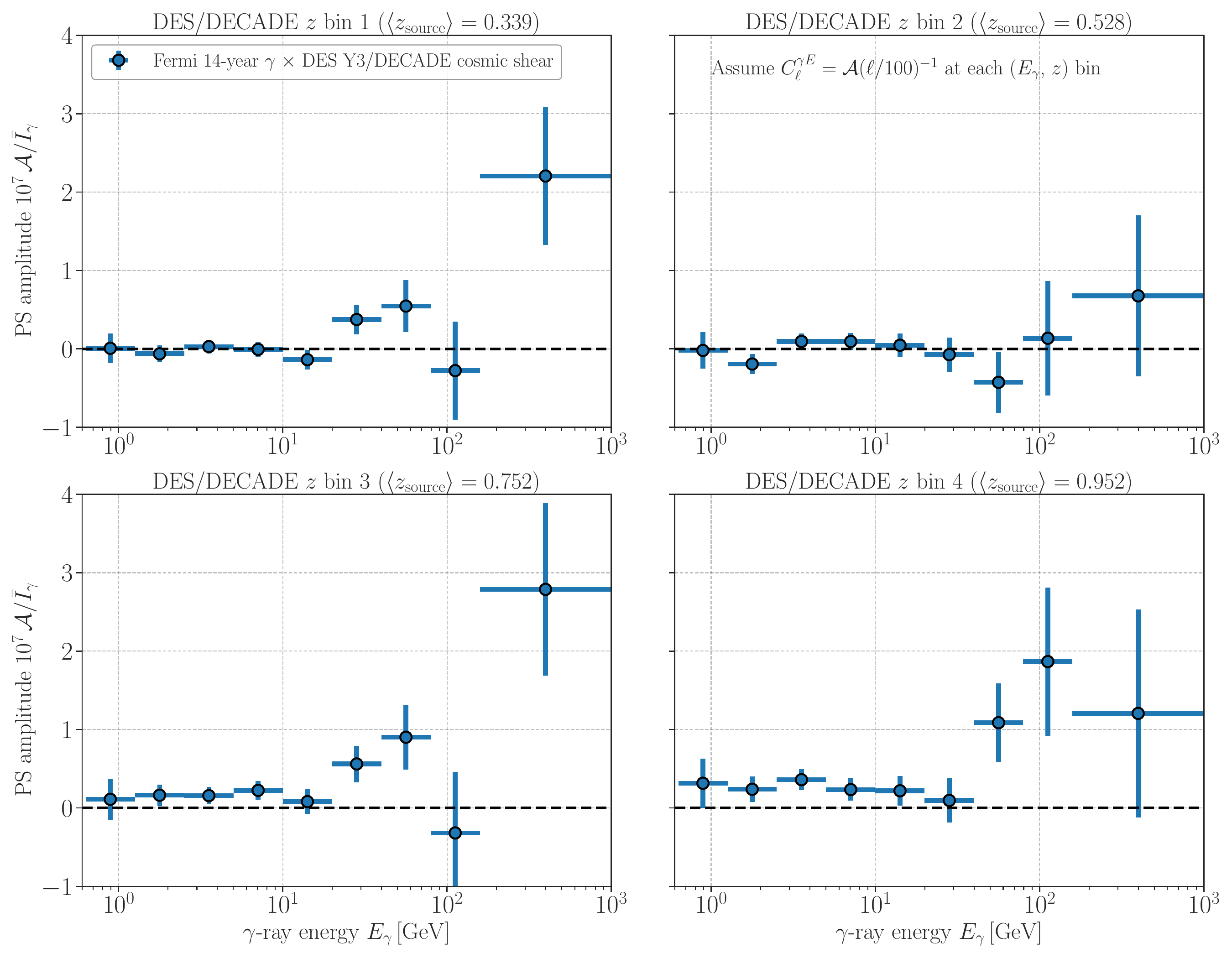}
\cprotect\caption{\label{fig:ps_amp_powerlaw} 
Dependence of the cross power spectra on $\gamma$-ray energy and source redshift.
Each panel shows the representative amplitude ${\cal A}$ normalized by the mean $\gamma$-ray intensity at the corresponding energy and redshift. See the main text for its definition.
The total signal-to-noise ratio over 9 energy and 4 redshift bins is 58.6, corresponding to a $p$-value of 0.01.
}
\end{figure*}

We also compare the 95\% upper bounds on $\langle \sigma v\rangle$ 
from large-scale cosmological cross-correlations for the $b\bar{b}$ channel with constraints from local structures, as shown in Fig.~\ref{fig:limit_DMann_comparisons}. 
While local probes such as the Galactic Center and dwarf spheroidal galaxies provide the tightest limits, cosmological analyses constrain DM on larger scales and using ensemble-averaged information and are less sensitive to object-to-object variations in the DM content.

For completeness, we also derive cosmological constraints on DM decay using our cross power spectra (Appendix~\ref{apdx:DMdecay}). In this case, the source field is the dark-matter overdensity, enabling more precise theoretical predictions. Our measurements place $2\sigma$ lower limits on the DM lifetime of $\sim 10^{26}$--$10^{27}\,\mathrm{s}$, depending on the decay channel.

\subsection{A model-dependent detection significance}\label{subsec:model_depend_s2n}

Our measurements are consistent with a null detection (Section~\ref{subsec:measurements}), based on a model-independent $\chi^2$ test. While conservative, this approach may not be optimal when reasonable signal priors exist. We therefore evaluate the detection significance using several template models to search for possible trends in the data.

As a consistency check with the DES-Y3 analysis \cite{DES:2025ulp}, 
we first test whether our data are described by the phenomenological models in Section~\ref{subsec:DESY3_model}. 
For a given template $C^{\gamma E}_\mathrm{model}(\ell)$, 
we perform a one-parameter likelihood analysis:
\beqa
\chi^2(q) &=& \sum_{\alpha, \beta, \alpha', \beta', i, j} 
\left[C^{\gamma E}_{\alpha \beta}(\ell_i)-q C^{\gamma E}_{\mathrm{model},\alpha\beta}(\ell_i)\right] \nonumber \\
&&
\times
\Gamma^{-1}_{ij}(\alpha\beta\alpha'\beta') \nonumber \\
&&
\times
\left[C^{\gamma E}_{\alpha' \beta'}(\ell_j)-q C^{\gamma E}_{\mathrm{model},\alpha'\beta'}(\ell_j)\right],
\eeqa
where $q$ is the amplitude parameter. The $1\sigma$ error satisfies $\chi^2(q)-\min(\chi^2)=1$.

Using the models in Section~\ref{subsec:DESY3_model}, we obtain
\beqa
q =
\begin{cases}
0.62 \pm 0.16 & \text{for the power-law model}, \\
0.079 \pm 0.071 & \text{for the log-parabola model},
\end{cases}
\eeqa
indicating a $3.9\sigma$ detection for the power-law model, while the log-parabola model is disfavored.

We further study the dependence on $\gamma$-ray energy and source redshift 
using a matched-filter approach. Following Ref.~\cite{DES:2025ulp}, 
the real-space correlation roughly scales as $\propto \theta^{-1}$, 
implying $C^{\gamma E}(\ell) \propto \ell^{-1}$. We fit
\beqa
\chi^2({\cal A}_{\alpha\beta}) &=& \sum_{\alpha, \beta, \alpha',\beta',i, j} 
\left[C^{\gamma E}_{\alpha \beta}(\ell_i)-{\cal A}_{\alpha\beta}\ell^{-1}_{i,100}
\right] \nonumber \\
&&
\times
\Gamma^{-1}_{ij}(\alpha\beta\alpha'\beta') \nonumber \\
&&
\times
\left[C^{\gamma E}_{\alpha'\beta'}(\ell_j)-{\cal A}_{\alpha'\beta'}\ell^{-1}_{j,100}\right],
\label{eq:chi2_ellinverse}
\eeqa
where $\ell_{100}=\ell/100$ and ${\cal A}_{\alpha\beta}$ is the amplitude in each bin.
The best-fit parameters can be obtained by the minimum of Eq.~(\ref{eq:chi2_ellinverse}),
while the covariance is
\beqa
\mathrm{Cov}\left[{\cal A}_{\alpha\beta}, {\cal A}_{\alpha'\beta'}\right]
= \left[\sum_{i,j}\Gamma^{-1}_{ij}(\alpha\beta\alpha'\beta')\ell^{-1}_{i,100}\ell^{-1}_{j,100}\right]^{-1}.
\eeqa
We then evaluate the detection significance as
\beqa
\left(\mathrm{S/N}\right)^2_{\cal A}
= \sum_{\alpha, \beta, \alpha', \beta'}\mathrm{Cov}^{-1}\left[{\cal A}_{\alpha\beta}, {\cal A}_{\alpha'\beta'}\right]
{\cal A}^{*}_{\alpha\beta} {\cal A}^{*}_{\alpha'\beta'},
\eeqa
where ${\cal A}^{*}_{\alpha\beta}$ is the best-fit parameters.
We find $(\mathrm{S/N})^2_{\cal A}=58.6$. With $9\times4=36$ degrees of freedom, this corresponds to a $p$-value of $\sim 0.01$.

Figure~\ref{fig:ps_amp_powerlaw} shows the energy and redshift dependence of ${\cal A}$, normalized by the mean $\gamma$-ray intensity. The amplitude follows a similar energy dependence to the mean $\gamma$-ray intensity, with mild deviations above 
$\sim100\,\mathrm{GeV}$. No strong energy-dependent features are found, supporting the power-law model in Ref.~\cite{DES:2025ulp}. The signal increases with source redshift, suggesting an origin linked to weak lensing by large-scale structure.

\section{Conclusions}\label{sec:conc}

We revisited the cross-correlation between the unresolved $\gamma$-ray background and galaxy shapes using 14 years of Fermi-LAT data and the DES Y3 and DECADE imaging surveys. The wide sky coverage ($\sim 12{,}000\,\mathrm{deg}^2$) enabled a precise search for annihilating dark matter on cosmological scales.

We measured the cross-correlation in Fourier space, in contrast to the DES-Y3 study \cite{DES:2025ulp}, which reported a detection using a matched-filter approach. Our measurements across 9 energy bins and 4 redshift bins were consistent with null based on a conservative, model-independent $\chi^2$ test. Using this result, we updated cosmological constraints on the annihilation cross section for masses between 7 and $10{,}000$ GeV. Assuming a substructure boost factor of $\sim100$ in Milky Way-sized halos, we excluded the canonical thermal cross section $\langle \sigma v\rangle=3\times10^{-26}\,\mathrm{cm}^3/\mathrm{s}$ for $7$--$40\,\mathrm{GeV}$ DM in the $b\bar{b}$ and $\tau^{+}\tau^{-}$ channels. Our limits approached the parameter space relevant for the Galactic center excess \cite{DiMauro:2021qcf} and the Galactic halo signal \cite{Totani:2025fxx}. We also found tension with 2--3 TeV co-annihilating DM scenarios such as the wino, 
for boost factors $\simgt 30$.

We validated our measurements with a model-dependent (matched-filter) analysis similar to Ref.~\cite{DES:2025ulp}. With the phenomenological power-law model, we found a $3.9\sigma$ preference, while a simple $\ell^{-1}$ template yielded a non-zero amplitude with a $p$-value of 0.01. This demonstrated that the inferred significance depend on the assumed signal model. The matched-filter analysis favored a power-law energy dependence, with a mild deviation from the mean $\gamma$-ray intensity scaling above $\sim 100\,\mathrm{GeV}$.

A decade after the first measurement \cite{Shirasaki:2014noa}, cross-correlations between the $\gamma$-ray background and galaxy shapes have provided DM constraints that are comparable and complimentary to local $\gamma$-ray probes such as nearby dwarf galaxies. Future improvements will require joint analyses with multiple large-scale structure tracers \cite{Rubiola:2025scy, Thakore:2026cxz}, allowing separation of $\gamma$-ray sources through environmental dependence. Precision cross-correlation studies will continue to offer key insights into the nature of DM.

\begin{acknowledgments}
This research is supported in part by JSPS KAKENHI grant 
Nos.~24H00215 (MS), 24H00221 (MS and NY), and 23H04899 (SH). This work was supported by NSF Grant No PHY-2209420 (SH). 
This work was supported by
World Premier International Research Center Initiative
(WPI Initiative), MEXT, Japan.
MS also aknowledge research supports by JST BOOST, Japan Grant Number JPMJBY24D8.
Numerical analyses were [in part] carried out on analysis servers at the Center for Computational Astrophysics, National Astronomical Observatory of Japan.
OM acknowledges support from the U.S. National Science Foundation under Grant No.~2418730.

This project used public archival data from the Dark Energy Survey (DES). Funding for the DES Projects has been provided by the U.S. Department of Energy, the U.S. National Science Foundation, the Ministry of Science and Education of Spain, the Science and Technology FacilitiesCouncil of the United Kingdom, the Higher Education Funding Council for England, the National Center for Supercomputing Applications at the University of Illinois at Urbana-Champaign, the Kavli Institute of Cosmological Physics at the University of Chicago, the Center for Cosmology and Astro-Particle Physics at the Ohio State University, the Mitchell Institute for Fundamental Physics and Astronomy at Texas A\&M University, Financiadora de Estudos e Projetos, Funda{\c c}{\~a}o Carlos Chagas Filho de Amparo {\`a} Pesquisa do Estado do Rio de Janeiro, Conselho Nacional de Desenvolvimento Cient{\'i}fico e Tecnol{\'o}gico and the Minist{\'e}rio da Ci{\^e}ncia, Tecnologia e Inova{\c c}{\~a}o, the Deutsche Forschungsgemeinschaft, and the Collaborating Institutions in the Dark Energy Survey.

The Collaborating Institutions are Argonne National Laboratory, the University of California at Santa Cruz, the University of Cambridge, Centro de Investigaciones Energ{\'e}ticas, Medioambientales y Tecnol{\'o}gicas-Madrid, the University of Chicago, University College London, the DES-Brazil Consortium, the University of Edinburgh, the Eidgen{\"o}ssische Technische Hochschule (ETH) Z{\"u}rich,  Fermi National Accelerator Laboratory, the University of Illinois at Urbana-Champaign, the Institut de Ci{\`e}ncies de l'Espai (IEEC/CSIC), the Institut de F{\'i}sica d'Altes Energies, Lawrence Berkeley National Laboratory, the Ludwig-Maximilians Universit{\"a}t M{\"u}nchen and the associated Excellence Cluster Universe, the University of Michigan, the National Optical Astronomy Observatory, the University of Nottingham, The Ohio State University, the OzDES Membership Consortium, the University of Pennsylvania, the University of Portsmouth, SLAC National Accelerator Laboratory, Stanford University, the University of Sussex, and Texas A\&M University.

Based in part on observations at Cerro Tololo Inter-American Observatory, National Optical Astronomy Observatory, which is operated by the Association of Universities for Research in Astronomy (AURA) under a cooperative agreement with the National Science Foundation.
\end{acknowledgments}

\appendix

\begin{figure*}[!t]
\includegraphics[clip, width=2.\columnwidth]{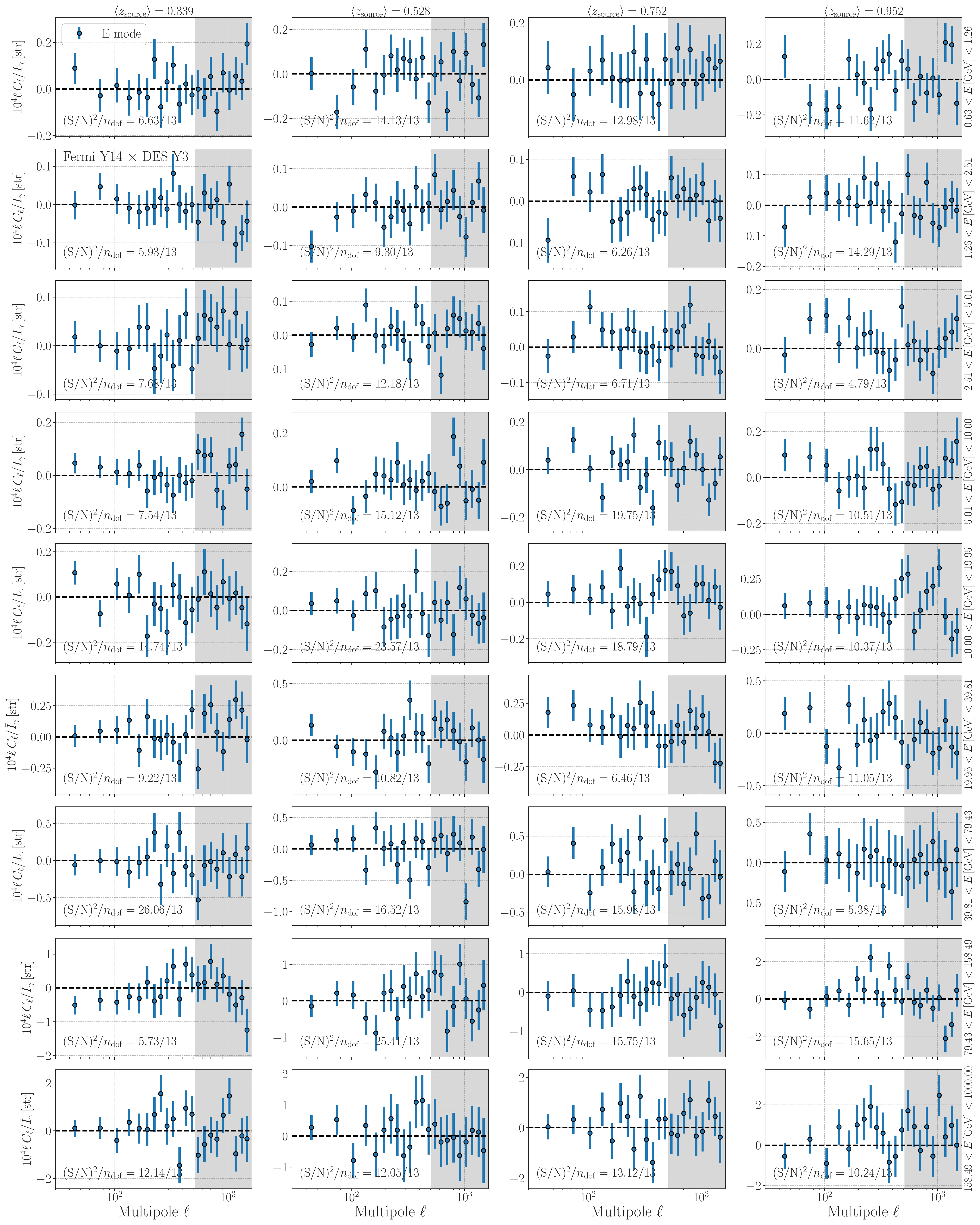}
\cprotect\caption{\label{fig:cl_desy3} 
Power spectra at individual $\gamma$-ray energy and galaxy redshift bins.
The average source redshift increases from left to right, while the $\gamma$-ray energy increases from top to bottom.
Data points in the gray region are excluded when computing the signal-to-noise ratio to ensure robust Fourier-space analysis.
This figure shows the measurements based on DES Y3 data.
}
\end{figure*}

\section{Power spectra at individual bins}\label{apdx:individual_cl}

\begin{figure*}[!t]
\includegraphics[clip, width=2.\columnwidth]{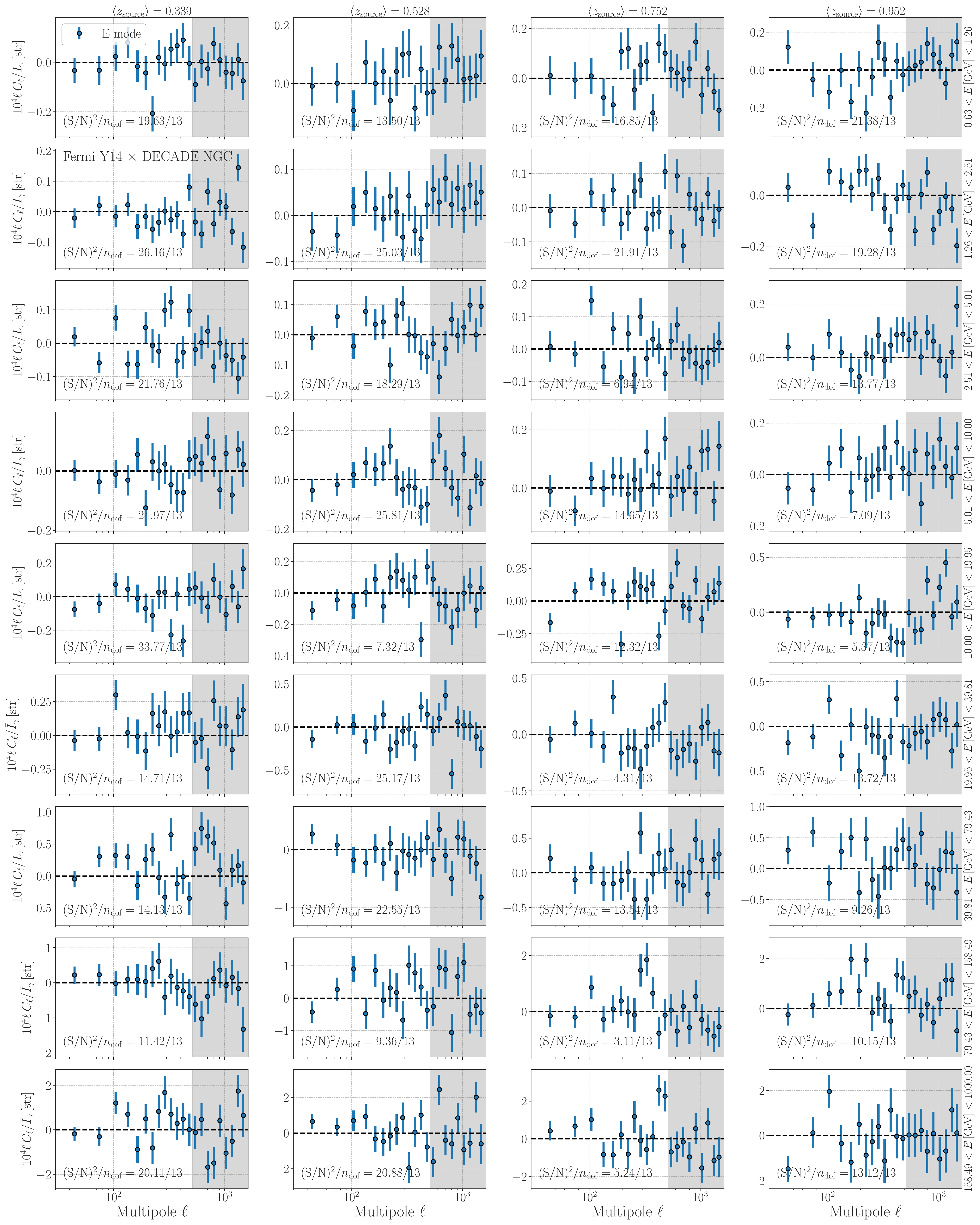}
\cprotect\caption{\label{fig:cl_decadengc} 
Similar to Fig.~\ref{fig:cl_desy3}, but showing the measurements based on DECADE NGC data.
}
\end{figure*}

\begin{figure*}[!t]
\includegraphics[clip, width=2.\columnwidth]{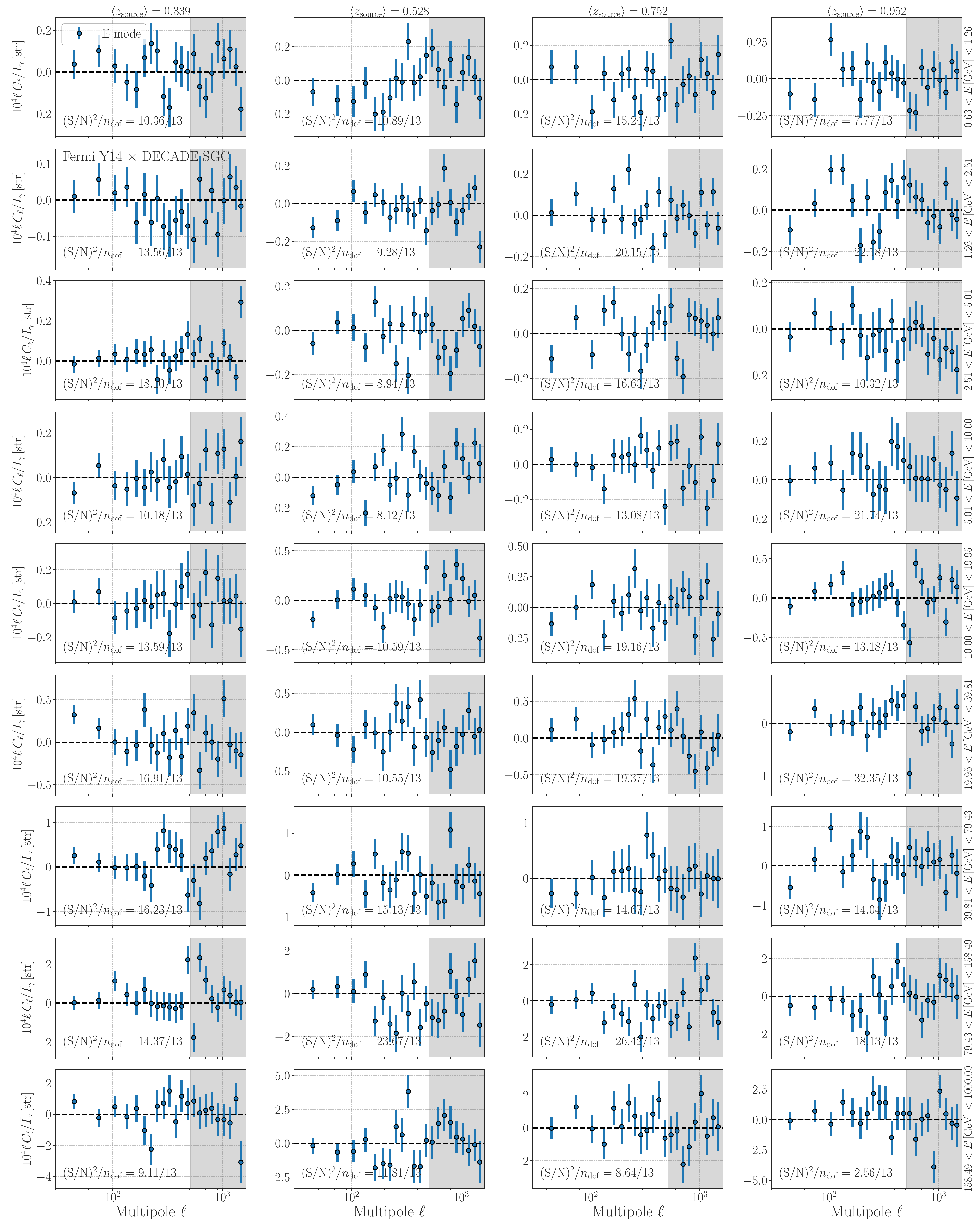}
\cprotect\caption{\label{fig:cl_decadesgc} 
Similar to Fig.~\ref{fig:cl_desy3}, but showing the measurements based on DECADE SGC data.
}
\end{figure*}

In this appendix, we present all measurements of the cross power spectra between the unresolved $\gamma$-ray background and galaxy shapes from DES Y3 and DECADE.  
Figures~\ref{fig:cl_desy3}, \ref{fig:cl_decadengc}, and \ref{fig:cl_decadesgc} show the E-mode spectra and their signal-to-noise ratios. No significant detections are found.

\section{Constraints of decaying DM}\label{apdx:DMdecay}

For decaying dark matter, the source function ${\cal S}$ and density field ${\cal F}$ in Eq.~(\ref{eq:Emissivity}) are
\beqa
{\cal S}(E_\gamma, z) &=&
\frac{\Gamma_{\rm d}}{m_{\rm dm}}
\left(\frac{{\rm d}N_{\gamma, d}}{{\rm d}E_\gamma}+Q_{{\rm IC}, d}(E_\gamma, z)
\label{eq:source_func_d}
\right) , \\
{\cal F}(\chi\bfn, z) &=& \rho_{\rm dm}(\chi\bfn, z),
\eeqa
where ${\rm d}N_{\gamma, d}/{\rm d}E_\gamma$ is the $\gamma$-ray spectrum per decay, $Q_{{\rm IC}, d}$ is the inverse-Compton contribution from decay-produced $e^{\pm}$, and $\Gamma_{\rm d}$ is the decay rate. We compute ${\rm d}N_{\gamma, d}/{\rm d}E_\gamma$ using the annihilation spectrum with a particle mass of $m_{\rm dm}/2$.
For the inverse-Compton contribution, we follow the method in Refs.~\cite{Ando:2015qda, Ando:2016ang}.

In Eq.~(\ref{eq:cross_powerspec}), we set $P_{\mathrm{m}{\cal F}}(k, z) = \bar{\rho}_{\rm dm}(z) P_{\delta}(k, z)$, where $P_{\delta}(k, z)$ is the matter power spectrum. We adopt the fitting formula of $P_{\delta}(k, z)$ calibrated against N-body simulations, which accurately captures non-linear effects on scales $\simgt 1\, {\rm Mpc}$ \cite{Takahashi:2012em}.

\begin{figure*}[!t]
\includegraphics[clip, width=2.\columnwidth]{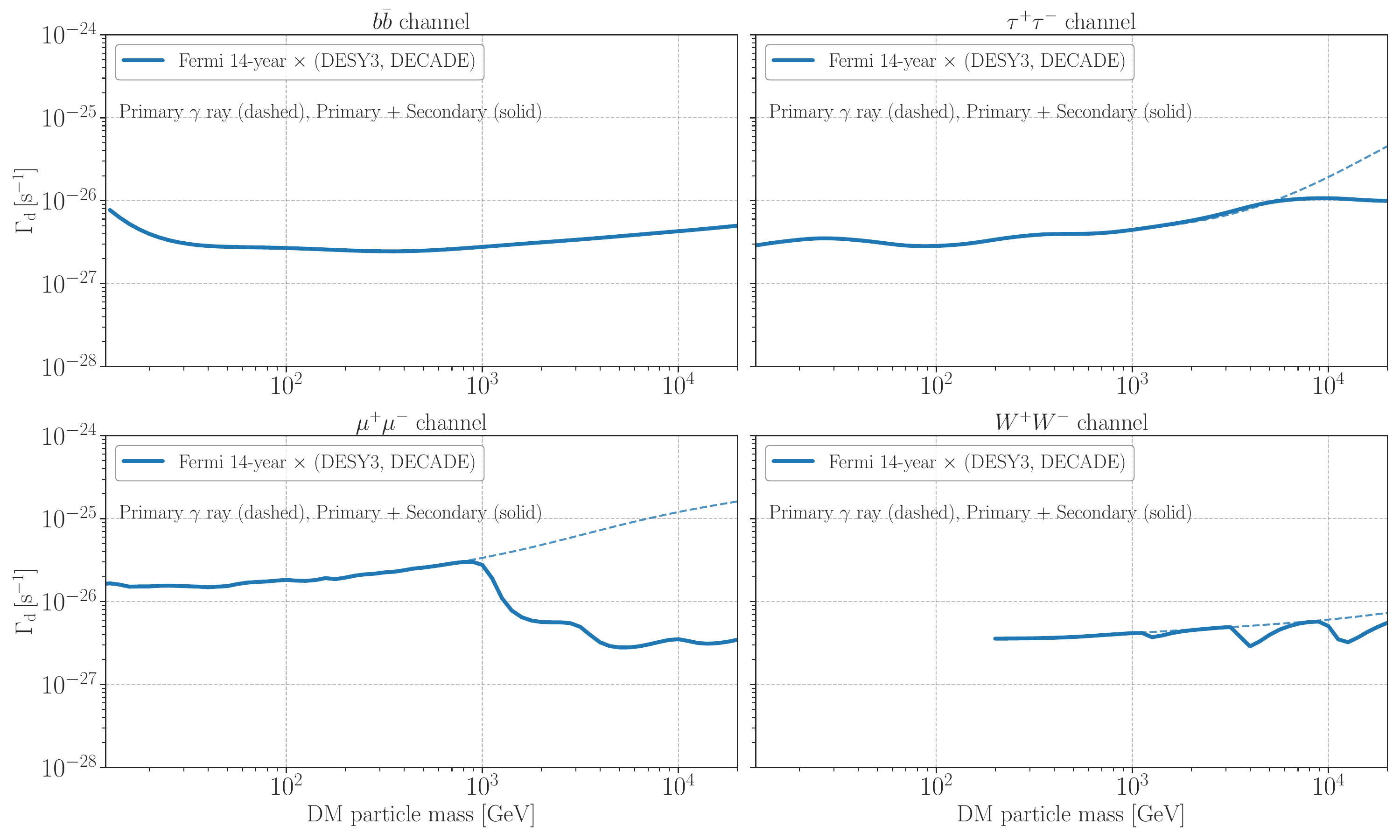}
\cprotect\caption{\label{fig:limit_DMdecay} 
95\% confidence upper limits on the DM decay rate as a function of DM mass. The four panels show the $b\bar{b}$, $\tau^{+}\tau^{-}$, $\mu^{+}\mu^{-}$, and $W^{+}W^{-}$ channels. Sold (dashed) lines include (exclude) secondary $\gamma$ rays from inverse-Compton scattering. 
}
\end{figure*}

Figure~\ref{fig:limit_DMdecay} shows the 95\% constraints on $\Gamma_{\rm d}$ obtained using the same likelihood analysis as in Section~\ref{subsec:DMann}. Our results place strong limits on decaying dark matter. Over a wide mass range, we find $\Gamma_{\rm d} \lesssim 10^{-26}$--$10^{-27}\,\mathrm{s}^{-1}$, comparable to constraints from galaxy-based cross correlations \cite{Paopiamsap:2023uuo} and the mean extragalactic $\gamma$-ray intensity \cite{Ibarra:2013cra}.

\bibliography{ref}

\end{document}